\def\ts     {\thinspace}
\def\kms    {\ifmmode{{\rm \ts km\ts s}^{-1}}\else{\ts km\ts s$^{-1}$}\fi}
\def\msol   {\ifmmode{{\rm M}_{\odot} }\else{M$_{\odot}$}\fi}
\def\lsol   {\ifmmode{L_{\odot}}\else{$L_{\odot}$}\fi}
\def\lfir   {\ifmmode{L_{\rm FIR}}\else{$L_{\rm FIR}$}\fi}
\def\,{\thinspace}
\def \ppm{$\pm$}

\documentclass[preprint]{aastex}
\usepackage{graphicx}
\usepackage{longtable}

 \received{} \accepted{} \paperid{}
\journalid{}{2007}

\begin{document}

\title{Molecular Gas and Star formation in ARP~302}
\author{Yiping Ao\altaffilmark{1,2},
Dinh-V-Trung\altaffilmark{3,4}, Jeremy Lim\altaffilmark{3}, Ji
Yang\altaffilmark{1} and Satoki Matsushita\altaffilmark{3}}

\altaffiltext{1}{Purple Mountain Observatory,
    Chinese Academy of Sciences, Nanjing 210008, China}

\altaffiltext{2}{Visiting Scholar, Academia Sinica Institute of
Astronomy and Astrophysics}

\altaffiltext{3}{Academia Sinica Institute of Astronomy and
Astrophysics, P.O. Box 23-141, Taipei 106, Taiwan}

\altaffiltext{4}{on leave from Institute of Physics, Vietnamese Academy of Science
and Technology, ThuLe, BaDinh, Hanoi, Vietnam.}

\begin{abstract}
We present the Submillimeter Array observation of the CO $J$=2$-$1
transition towards the northern galaxy, ARP~302N, of the early
merging system, ARP~302. Our high angular resolution observation
reveals the extended spatial distribution of the molecular gas in
ARP~302N. We find that the molecular gas has a very asymmetric
distribution with two strong concentrations on either side of the
center together with a weaker one offset by about 8 kpc to the
north. The molecular gas distribution is also found to be
consistent with that from the hot dust as traced by the 24 $\mu$m
continuum emission observed by the Spitzer. The line ratio of CO
$J$=2$-$1/1$-$0 is found to vary strongly from
about 0.7 near the galaxy center to 0.4 in the outer part of the
galaxy. Excitation analysis suggests that the gas density is low,
less than 10$^3$ cm$^{-3}$, over the entire galaxy. 
By fitting the SED of ARP~302N in the far infrared we obtain a dust 
temperature of $T\rm _d$=26--36 K and a dust mass of 
M$\rm _{dust}$=2.0--3.6$\times$10$^8$ M$\rm _\odot$.
The spectral index of the radio continuum is around 0.9. The
spatial distribution and spectral index of the radio continuum
emission suggests that most of the radio continuum emission is
synchrotron emission from the star forming regions at the nucleus
and ARP302N-cm. The good spatial correspondance between the 3.6 cm radio continuum
emission, the Spitzer 8 \& 24 $\mu$m data and the high resolution
CO $J$=2$-$1 observation from the SMA shows that there is the
asymmetrical star forming activities in ARP~302N.

\end{abstract}


\section{Introduction}

Most luminous infrared (IR) galaxies (LIRGs; 10$^{11}$ L$\rm
_\odot$ $<$ L$\rm _{IR}$ $\leq$ 10$^{12}$ L$\rm _\odot$) are found
in interacting or merging systems with disturbed morphologies. In
the standard scenario proposed by Sanders et al. (1988a,b), the
formation of LIRGs, or ultra-luminous infrared galaxies (ULIRGs)
is due to the interaction and merging of two gas-rich galaxies.
Numerical simulations of interacting galaxies
\citep{barnes91,mihos96} indicate that a stable bar is formed
during the first encounter. Under the gravitational torque induced
by the bar, the molecular gas is channelled into the central
regions of the galaxies forming a huge reservoir to fuel intense
starbursts. Recent high angular resolution observations \citep{downes98} seem to support
this picture by revealing rotating disks of molecular gas that has
been driven into the centers of the mergers.

A lot of (U)LIRGs have been studied at high angular resolution, for example 
ARP~299 \citep{casoli99}, NGC5394/95 \citep{kaufman02}, ARP~220 \citep{wiedner02}, 
NGC4038/39 \citep{zhu03}, NGC5194/95 \citep{matsushita04}, NGC~6090 \citep{wang04}, 
NGC5218 \citep{olsson07}. A few samples of the sources along the merging sequence 
are also been investigated. \citet{xilouris04} found evidence for an increase 
in 100- to 850-$\mu$m flux density ratio along the merging sequence and 
interpreted this trend as an increase of the mass fraction of the warm dust in 
comparison to the cold dust, inferring the possible enhanced star formation 
rate (SFR). \citet{bergvall03}, however, question the
ubiquitous connection between interactions and starburst activity.
\citet{gao99} find the enhanced star formation efficiency (SFE) for
the later merging systems, but the effect is cancelled by the decrease
in the amount of molecular gas. That might result in no change in the infrared luminosity, i.e.
the indicator of the starburst activity. Thus more sources in high
resolution are needed and multiple CO transtions are essential to
investigate the gas properties for the mergers at different
stages.

Here we present another case study of an early merger system,
ARP~302. This galaxy is also named as VV 340 or UGC 9618, which consists of two galaxies
(ARP~302N and ARP~302S) with a separation of about 40$\arcsec$ between the two
nuclei, which is $\sim$27 kpc at the angular distance of 138 Mpc
(the luminosity distance is 147 Mpc), and has an infrared
luminosity of 4.5$\times$10$^{11}$ L$_\odot$. ARP~302 is the most
CO luminous system known in the local universe, and more luminous
in CO than those ultraluminous galaxies out to redshifts of 0.27
(Solomon et al. 1997). The detected molecular gas mass is
9.1$\times$10$^{10}$ M$_\odot$ \citep{lo97} (corrected by the adopted
comsmology here). HI data gives a mass
of atomic hydrogen 2.7$\times$10$^{10}$ M$_\odot$ \citep{van01}.
In all observations it was found that
ARP~302N contributes most of the IR emission and molecular gas
mass.

Previous high resolution CO $J$=1$-$0 observation on ARP~302 by
\cite{lo97} (the angular resolution is
7$\arcsec$$\times$5$\arcsec$) reveals two extended CO disks ($>$
10 kpc), and ARP~302N shows roughly an exponential disk with a
scale length of 7 kpc extending out to 23 kpc in diameter.
Based on the unusually extended distribution of the molecular gas,
\cite{lo97} concluded that the high IR luminosity in ARP~302
arises simply because an extremely large amount of molecular gas
is forming high-mass stars at a rate similar to the Galaxy, instead of the
starburst activities in the galactic nucleus.

In this study, we report the high angular
resolution CO $J$=2$-$1 observation of ARP~302N and the analysis of archival
data in the radio and mid/far infrared to study the physical properties of the gas
as well as the star formation activities in this early merger. In \S\ref{observations},
we show the details of our observations together with the
information about the archive data used in this paper. In
\S\ref{spatial}, \S\ref{ratio} and \S\ref{kinematics}, we present
the results of spatial distribution and the kinematics of the
molecular gas of our observations. In \S\ref{vla} and
\S\ref{spitzer}, the results of the radio emission and the mid/far
infrared emission from the archive data are presented. Finally we
present the dust properties using a dust model in \S\ref{dust},
the excitation condition of the gas in \S\ref{excitation}, the
star formation activities in \S\ref{formation} and a comparison with
other mergers in \S\ref{compar}. A summary of this
study is given in \S\ref{conclusion}. In this paper, we use a
$\Lambda$ cosmology with $\rm H_{\rm0} =
71$~km~s$^{-1}$~Mpc$^{-1}$, $\rm \Omega_\Lambda=0.73$ and $\rm
\Omega_{m}=0.27$ \citep{spergel03}.

\section{Observations}\label{observations}

\subsection{CO $J$=2$-$1 observation with the Submillimeter array}

We carried out the observation of the CO $J$=2$-$1 line from
ARP~302N on 2005 March 26, using the Submillimeter array
(SMA\footnotemark[5])\footnotetext[5]{The Submillimeter Array is a
joint project between the Smithsonian Astrophysical Observatory
and the Academia Sinica Institute of Astronomy and Astrophysics,
and is funded by the Smithsonian Institution and the Academia
Sinica.} \citep{ho04} in its most compact configuration. At that
time the SMA consists of 7 antennas of 6 meter in diameter. The
weather condition during our observations is average, with
atmospheric opacity at zenith $\tau_{225}$ $\sim$ 0.35. The single
sideband system temperature is about 450 K. Because of the large
extent of ARP~302N, we cover the whole galaxy using a two-pointing
mosaic, with a separation of 20$\arcsec$ between the two pointings
located at 10$\arcsec$ north and south of the nucleus
($\alpha$=14$^h$57$^m$00$^s$.67, $\delta$=+24$\rm
^o$37$\arcmin$02$\arcsec$.8).

The total on-source time was about 2 hours per one pointing. We
observed the CO $J$=2$-$1 line in the upper sideband of the double
sideband receiver. The SMA digital correlator was configured to
provide twenty-four independent spectral windows in each sideband.
Each window was divided into 128 channels, which provided 0.8 MHz
frequency resolution (i.e. $\sim$1.1~km~s$^{-1}$) at 223 GHz,
giving a total frequency coverage in each sideband of 1.97 GHz
(i.e. $\sim$2650~km~s$^{-1}$). We tuned the SMA receivers to the
redshifted CO $J$=2$-$1 line at the frequency of 222.967 GHz,
which corresponds to the systemic velocity of 10180
\kms. We observed Jupiter and Callisto for bandpass and flux
calibration respectively, and estimated the flux density scale to
be accurate to about 10\% at 1.3 mm band. The nearby quasar
1357+193 was observed every 20 minutes to correct for
time-dependent gain variations due to atmospheric fluctuations.
The data were calibrated using the MIR-IDL package, which was
specially developed for the SMA. We then exported the data into
MIRIAD format for further processing. The mosaiced images made
from SMA data were deconvolved using the Steer-Dewdney-Ito clean
algorithm \citep{steer84}. The synthesized beam size is
3.4$\arcsec$$\times$2.5$\arcsec$ for the CO $J$=2$-$1 observatrion
with the uniformed weighting and the resulting noise level is 60
mJy beam$^{-1}$ at a velocity resolution of 20 km~s$^{-1}$. The 230
GHz continuum image of ARP~302N was formed from the line-free
channels and the rms noise level is 5 mJy~beam$^{-1}$. 
No 1.3 mm continuum emission was detected from our
observation at 3 $\sigma$ level of 15 mJy~beam$^{-1}$.

\subsection{CO $J$=2$-$1 observation with the Submillimeter Telescope}
To estimate the amount of CO J=2--1 flux recovered in our SMA
observations, we also observed the CO $J$=2$-$1 line in ARP~302N
with a single dish telescope, the Submillimeter Telescope (SMT),
in Arizona on 2005 December 14 under excellent weather condition
($\rm \tau_{225 GHz}$ $\sim$ 0.08). The observation was done in
beam switching mode, with a beam throw of 2 arcmins and a chopping
frequency of 1.25 Hz. The half-power beamwidth (HPBW) of the SMT
telescope at the frequency of CO J=2--1 line is 28$\arcsec$, which
is large enough to cover most of the CO emitting region of
ARP~302N. The acousto-optical spectrometer (AOS) was used as the
backend, with a usable bandwidth of 1 GHz and a velocity
resolution of 1.3~km~s$^{-1}$. A single pointing observation was
made toward the nucleus of ARP~302N. The on-source integration
time was 45 minutes. The data was reduced with the GILDAS software
package. All the spectra were then co-added to produce the final
CO J=2--1 spectrum of ARP~302N. The final spectra were smoothed to
the resolution of 14 km~s$^{-1}$. The
noise level on antenna temperature scale was 3 mK. We
converted the spectrum into main beam temperature scale using a
main beam efficiency of 0.66.

\subsection{VLA archive data}
We searched the Very Large Array data archive, which is provided by the National Radio Astronomy
Observatory
(VLA\footnotemark[6]\footnotetext[6]{The National Radio Astronomy
Observatory is a facility of the National Science Foundation
operated under cooperative agreement by Associated Universities,
Inc.}), for data on ARP~302N. Previous observations were found for
X band (8.44 GHz, i.e. 3.6~cm), C band (4.86 GHz, i.e. 6.0~cm) and
L band (1.49 GHz, i.e. 20~cm). We used the NRAO data reduction
package to reduce the available data sets. Continuum images were
made using a natural weighting. For X-band data, which were obtained in
the extended configuration, we
applied tappering to the data in order to recover the more extended continuum emission.

\subsection{Spitzer archive data}
To compare with our SMA data, we also searched for the archival
data on ARP~302N produced by the Spitzer telescope. We found that
ARP~302N was imaged with the Infrared Array Camera (IRAC; Fazio et
al. 2004) in all four channels (3.6, 4.5, 5.8, and 8.0 $\mu$m) on
2005 July 16 and with Multiband Imaging Photometer for Spitzer
(MIPS; Rieke et al. 2004) in 24, 70 and 160 $\mu$m on 2005 January
25. The images were reduced with the Spitzer Science Center (SSC)
IRAC reduction pipeline and combined with the SSC mosaicker. The
point spread functions are
2.2$\arcsec$, 6$\arcsec$, 18$\arcsec$ and 40$\arcsec$ for 8, 24,
70 and 160 $\mu$m wavelengths, respectively. We only use the 8.0 $\mu$m and
the MIPS data in this paper. The uncertainties on the final
absolute calibrations are estimated at 10\%, 10\%, 20\%, and 20\%
for the 8, 24, 70, and 160 $\mu$m data, respectively.

\section{Results}\label{results}

\subsection{Spatial distribution of the molecular gas}\label{spatial}

\begin{figure}
 \includegraphics[angle=-90,width=0.75\textwidth]{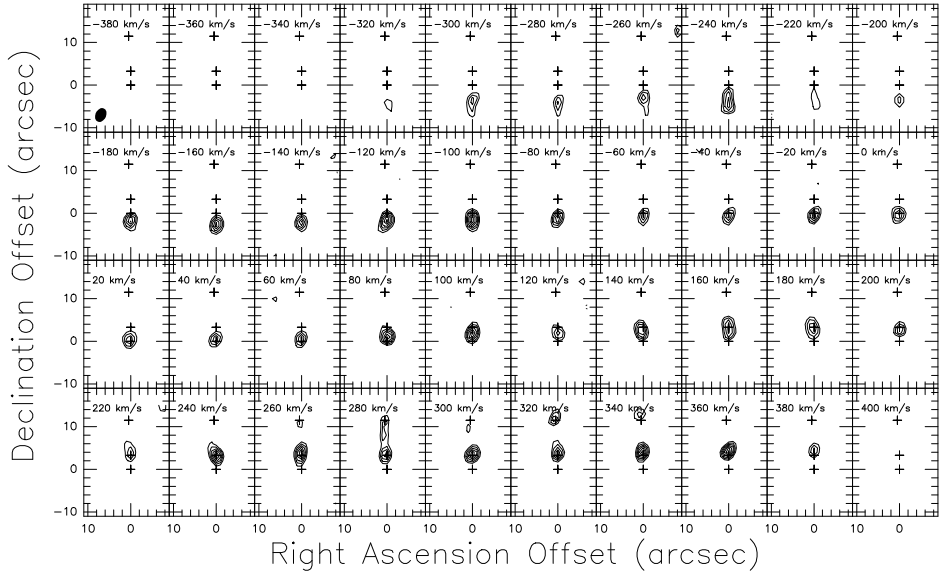}\\
  \caption{Channel maps of CO $J$=2$-$1 line of ARP~302N. The contours are -3, 3, 4, 5, 6, 7 and 8 $\times$ 60 mJy~beam$^{-1}$
  (1 sigma level), with a synthesized beam of 3.4$\arcsec$$\times$2.5$\arcsec$.
  Three cross symbols from bottom to top show the positions of nuclear
  region, ARP302N-cm (see \S\ref{vla}) and outmost northern CO $J$=2$-$1 emission peak respectively.
  The map is centered on the nucleus of ARP~302N ($\alpha(\rm J2000)$=14$\rm^h$57$\rm^m$00$\rm^s$.67,
  $\delta(\rm J2000)$=+24$\rm ^o$37$\arcmin$02$\arcsec$.8).}\label{channel_map}
\end{figure}

In Figure~\ref{channel_map} we show the channel maps of CO
$J$=2$-$1 emission for ARP~302N. There is a clear positional shift
of the emission centroid with the velocity along the north-south
direction, consistent with the rotation of the gas in the galactic
disk. In each velocity channel we find that the emission is just
barely resolved in the east-west direction, which is perpendicular
to the galactic plane. In the channels with the velocity within
\ppm 80 \kms, the emission is compact and only slightly resolved
by the SMA, showing a symmetric distribution. The emission in this
velocity range is centered at the nucleus and the emission of some
channels is weaker than the channels outside this range. In higher
redshifted and blueshifted velocity channels the emission is
slightly more extended in the north-south direction, especially
for the channels above \ppm 220 \kms. For the receding part of the
galaxy the CO J=2--1 emission is resolved into two distinct
components, which are separated by about 8$\arcsec$.

\begin{figure}
  \includegraphics[angle=0,width=0.5\textwidth]{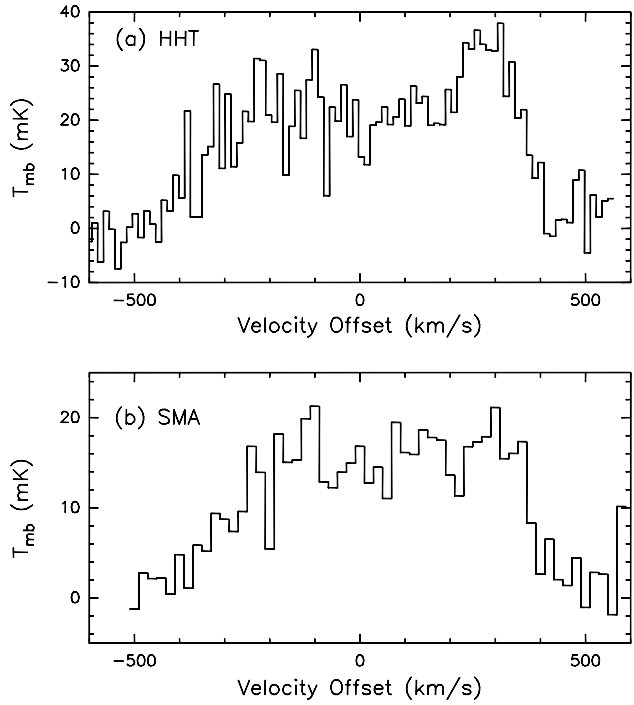}\\
  \caption{ (a).  At the position of the nucleus of ARP~302N,
  spectrum of CO $J$=2$-$1 line obtained with the Sub-millimeter Telescope 
  at a resolution of 28$\arcsec$. The measurement is on the main beam
  brightness temperature. (b). Spectrum of CO $J$=2$-$1 line at the
  same position obtained from the SMA convolved to the same 
  angular resolution as the single dish observation.}\label{spectra}
\end{figure}

Because the total extent of the CO emission is significant by
comparison with the primary beam of the SMA, part of the CO flux
could be resolved out in our SMA observations. In
Figure~\ref{spectra} we compare the spectrum of CO J=2--1 obtained
with the SMT at the resolution of 28$\arcsec$ and that obtained by
convolving our SMA channel maps to the same resolution. The CO
$J$=2$-$1 flux detected by the SMA within one SMT beam centered at
the nucleus is 356($\pm$8) Jy km~s$^{-1}$. It suggests that the SMA
recovers about 60\% of the CO $J$=2$-$1 flux from Arp 302 in
comparison with a total flux of 576 Jy \kms detected by the SMT. The
total CO $J$=2$-$1 flux detected by the SMA is 382 ($\pm$10) Jy
km~s$^{-1}$. The most significant difference between both spectrum
is the comparatively lower emission in the SMA spectrum from 220 to
320 \kms, which could be due to extended emission that is partially
resolved out by the SMA in the outer northern region. The similarity
in the line shape between single dish and the SMA spectrum suggests
that the missing flux is more or less uniformly distributed across
the line.

\begin{figure}
  \includegraphics[angle=0,width=0.95\textwidth]{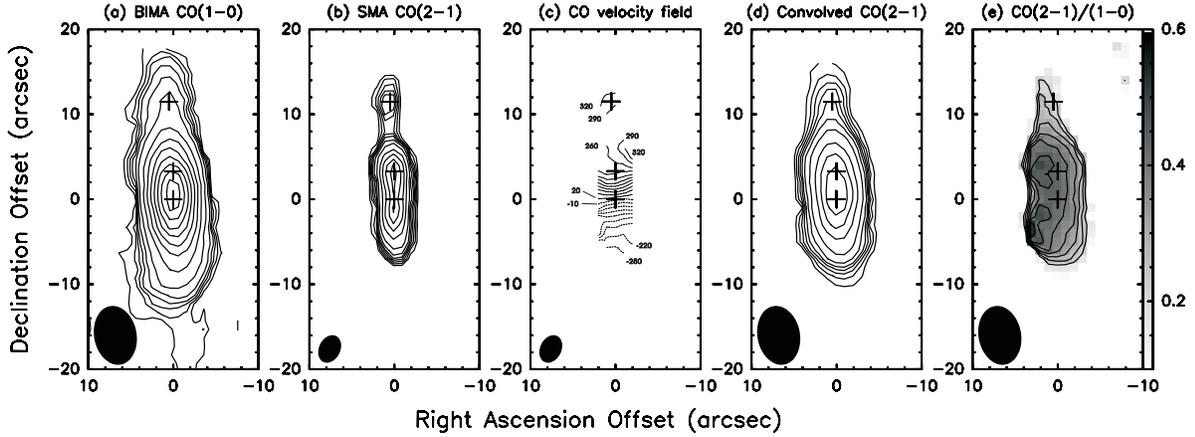}\\
  \caption{(a). Integrated intensity map of CO $J$=1$-$0 line of ARP~302N.
  CO data is from BIMA observation, obtained by Lo et al 1997. The contours shown are
  6, 9, 12, 15, 20, 30, 50, 70, 90, 110, 130, 150 and 170 $\times$ 0.6 Jy km~s$^{-1}$.
  (b). Integrated intensity map of CO $J$=2$-$1 line of ARP~302N.
  The contours shown are 3, 6, 9, 12, 15, 20, 30, 50, 70, 90, 110 and 130 $\times$ 1 Jy km~s$^{-1}$.
  (c). CO $J$=2$-$1 velocity field obtained by the SMA. Contour levels are from -250 to 320 \kms\,
   by the step of 30 \kms\,. Dashed line is blueshifted velocity and solid line redshifted.
  (d). CO $J$=2$-$1 integrated intensity map convolved to the BIMA resolution.
  The contours shown are 3, 6, 9, 12, 15, 20, 30, 50, 70 and 90 $\times$ 2 Jy km~s$^{-1}$.
  (e). Ratio map of $T\rm_{CO J=2--1}$/$T\rm_{CO J=1--0}$ of ARP~302N.
  The contour is from 0.2 to 0.45 by the step of 0.05.
  The cross symbols are the same as in Figure~\ref{channel_map}.
  }\label{arp302_color}
\end{figure}

Figure~\ref{arp302_color}(b) shows the contours of integrated CO
$J$=2$-$1 intensity map at 3.4$\arcsec$$\times$2.5$\arcsec$
resolution by the SMA. The molecular gas almost covers the entire
optical disk of the galaxy seen almost edge on and distributes
over a region with the size of about 15 kpc. We find that the
molecular gas has a very asymmetric distribution with two strong
concentrations on either side of center together with a outmost
weaker concentration offset by about 8 kpc to the north. In the
figure, we show three crosses from bottom to top: one is at the
nucleus, the second is at the peak of 3.6 cm emission outside of
the nucleus (see \S\ref{vla}) and the third one is at the outmost
north for the weak CO J=2--1 emission concentration. To compare
with the BIMA data, we show the BIMA CO $J$=1$-$0 map in
Figure~\ref{arp302_color}(a). Our higher resolution map allows us
to investigate the molecular gas distribution with three
concentrations along the north-south direction, which the BIMA
data can not tell us. We also convolved the SMA beam to the BIMA's
and the map is shown in Figure~\ref{arp302_color}(d).
The BIMA map shows more extended distribution of the gas with large
radial distance comparison with the SMA result, which may be
due to the different exciation conditions for the gas at large
radial distance.

\subsection{CO $J$=2$-$1/$J$=1$-$0 line ratio}\label{ratio}

In Figure~\ref{arp302_color}(e), we present the map of the brightness temperature ratio of CO
$J$=2$-$1 emission from the SMA to CO $J$=1$-$0 emission from the
BIMA. To calculate the line ration, we first convolve our CO $J$=2$-$1 map to the same
angular resolution of 7$\arcsec$$\times$5$\arcsec$ of the BIMA
data as shown in Figure~\ref{arp302_color}(d) and form the line ratio map. We find that the line ratio varies from
0.2 at the outer region to about 0.5 at the peaks. The ratio
should be the lower limit since the missing flux of the SMA is
about 36\% but there is no missing flux for the BIMA data -- the
flux of 304$\pm$45 Jy km~s$^{-1}$ from the BIMA CO $J$=1$-$0
observation \citep{lo97} is similar to the value of 290$\pm$35 Jy
km~s$^{-1}$ from NRAO 12m telescope (private communication from Y.
Gao; Gao \& Solomon 1999). The line ratio is a very important observational
constraint to estimate the physical conditions of the molecular gas in 
Arp~302N as we will show in the next section.

\subsection{Kinematics}\label{kinematics}

\begin{figure}
  \includegraphics[angle=0,width=0.5\textwidth]{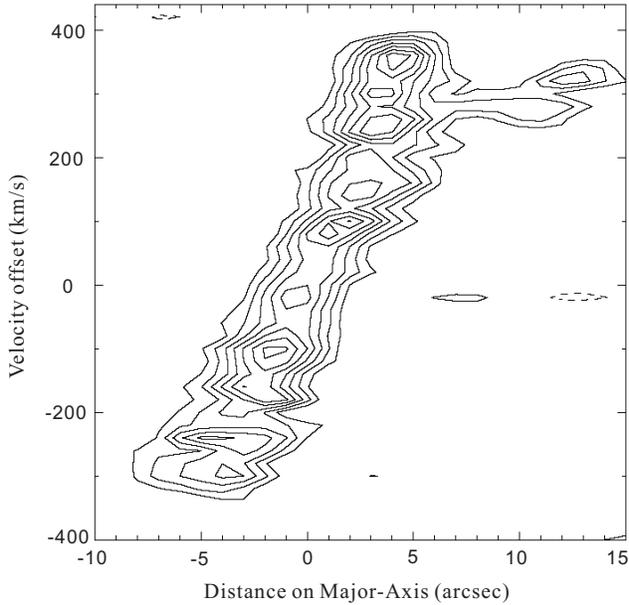}\\
  \caption{Position-velocity diagram of ARP~302N along major axis. The contour levels are
  from 0.12 to 0.54 Jy~beam$^{-1}$ by a step of 0.06 Jy~beam$^{-1}$.}\label{velplot}
\end{figure}

We present in Figure~\ref{arp302_color}(c) the intensity-weighted velocity 
field as contours. A velocity field typical of normal rotation is clearly 
seen along the major axis within a radius of $\sim$4$\arcsec$, i.e. 2.7~kpc.
Along the minor axis, we can not detect any disturblance of the gas in the inner
region. In Figure~\ref{velplot} we shows the position-velocity (PV) 
diagram of the CO J=2--1 emission along the major axis (position 
angle of 1$^\circ$). We note that the molecular gas exhibits a
nearly constant velocity gradient (about 100 km~s$^{-1}$
kpc$^{-1}$) within the central 6$\arcsec$ or 4.0 kpc in linear scale. 
In the velocity range
between 260 and 340 km~s$^{-1}$ there are clearly two components: one is associated
with the gas at ARP302N-cm and the other component located to the
northern part about 8$\arcsec$ to 15$\arcsec$ away from
the nucleus. The gas at 3.3$\arcsec$ (2.2 kpc) north of the
nucleus shows a slightly steeper velocity gradient (about 170
km~s$^{-1}$ kpc$^{-1}$) and the position of this feature is
consistent with ARP302N-cm (the secondary 3.6 cm radio emission
feature, see \S\ref{vla}). The molecular gas located at large
galactocentric distance toward the north shows clear deviation from
solid body rotation, as can be seen in Figure~\ref{velplot}.

\cite{Sakamoto99} find that there is a steep rise in rotation
velocity, i.e. the velocity gradient $\sim$ 1000 km s$^{-1}$
kpc$^{-1}$, in the central regions with radii of typically
$\leq$ 0.5 kpc towards most nearby spiral galaxies. 
It is therefore apparent that the presence of a shallow velocity
gradient over an extended region of 4.0 kpc in radius sets Arp~302N apart from
normal cases. However, solid body
rotation over kpc-scales is found in some special cases, such as NGC~4194 \citep{aalto00}, 
NGC~7479 \citep{laine99}, UGC~2866 \citep{huettem99} and NGC~891 \citep{garcia95}.
Usually the presence of a molecular ring
or the solid body rotation in bars are evoked to explain the observed linear velocity gradient.
\subsection{VLA results}\label{vla}
\begin{figure}
  \includegraphics[angle=0,width=0.5\textwidth]{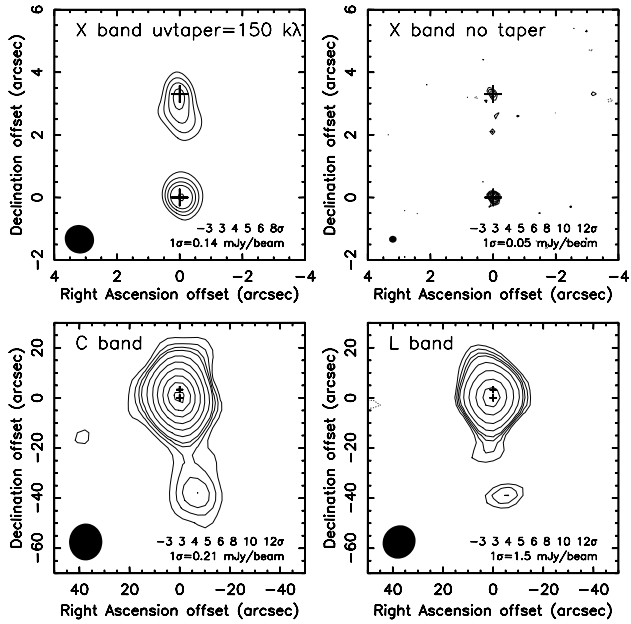}\\
  \caption{Radio images are shown towards ARP~302.
  The images at 3.6~cm are shown at two top panels with the resolutions of 0.76$\times$0.54
  (left top panel) and 0.23$\times$0.09 (right top panel).
  The images at 20~cm and 6~cm are shown at bottom panels with resolutions of
  13.7$\arcsec$$\times$12.7$\arcsec$ (left bottom panel) and
  14.8$\arcsec$$\times$13.3$\arcsec$ (right bottom panel).
  The noise values and contour levels are shown at the right
  bottom of each panel.
  The cross symbols are the same as in Figure~\ref{channel_map}.}\label{arp302_vla}
\end{figure}

Low resolution of the images at C band (6~cm) and L band (20~cm),
as shown in Figure~\ref{arp302_vla}, indicate that
the radio emission is concentrated in the central region. There is
only one central peak and the details can not be well resolved
with the current poor resolutions, which are 13--14$\arcsec$ for
both observations. The integrated intensities are 30.2$\pm$0.4 mJy
and 89.0$\pm$3.6 mJy for 6~cm and 20~cm respectively, yielding a
spectral index, $\alpha$ (where S $\varpropto$ $\nu^{-\alpha}$),
of 0.9.

High resolution of the X band (3.6~cm) images are shown in
Figure~\ref{arp302_vla}. 3.6 cm continuum emission is mainly
contributed from two components with similar total integrated
intensity. One component, associated with the galactic nucleus, is
compact and marginally resolved (right top panel of
Figure~\ref{arp302_vla}). A second component is located at
3.3$\arcsec$ ($\sim$ 2.2 kpc) north of the nucleus (ARP302N-cm
hereafter) and extends from 2$\arcsec$ to 4$\arcsec$ away from the
nucleus (left top panel of Figure~\ref{arp302_vla}). All the results
are summarized in Table~\ref{xband}.

\subsection{Spitzer results}\label{spitzer}

\begin{figure}
  \includegraphics[angle=-90,width=0.5\textwidth]{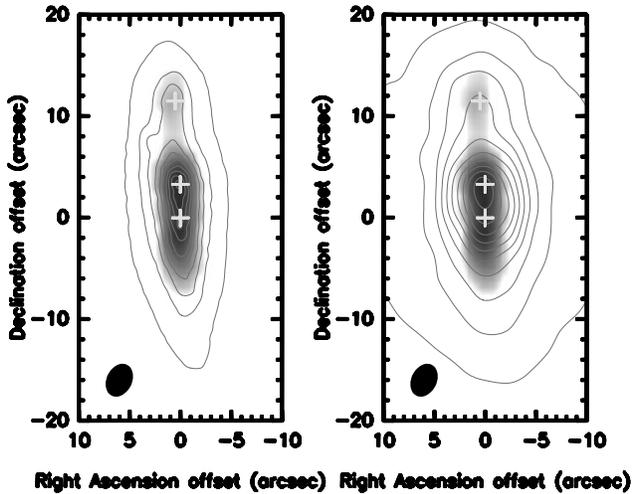}\\
  \caption{The Spitzer 8 $\mu$m and 24 $\mu$m images are shown in contours and overlaid
  on the SMA CO $J$=2$-$1 integrated intensity in grey scale.
  The contour levels are from 5\% to 95\% by a step of 10\% of the peak values.
  The cross symbols are the same as in Figure~\ref{channel_map}.}\label{irac}
\end{figure}

In Figure~\ref{irac} we show the mid infrared continuum emision from
Arp~302 at 8.0 $\mu$m and 24 $\mu$m taken with IRAC and MIPS instruments
of Spitzer. Both 8.0 $\mu$m and 24 $\mu$m images exhibit similar
morphology to that seen in the CO J=2--1 emission. 
Hot dust traced by the 8.0 $\mu$m and 24 $\mu$m share the
similar spatial distribution as the molecular gas traced by the CO
J=2--1 emission. With a resolution of $\sim$7$\arcsec$ at 24
$\mu$m, ARP~302 is very well resolved. From the MIPS image we estimate the flux densities
of 0.32$\pm0.03$ Jy and 0.08$\pm0.01$ Jy for ARP~302N and ARP~302S
respectively. The flux densities are 5.16$\pm1.03$ Jy at 70 $\mu$m
and 5.08$\pm1.02$ Jy at 160 $\mu$m for ARP~302N respectively.
Applying the same flux ratio based on 24~$\mu$m between ARP~302N
and ARP~302S to the total IRAS flux densities of ARP~302, we
estimate 5.32$\pm1.60$ Jy at 60 $\mu$m and 11.6$\pm3.5$ Jy at 100
$\mu$m for ARP~302N.

\section{Discussion}\label{discussion}

\subsection{Dust properties constrained by one component dust
model}\label{dust}

The dust continuum allows us to investigate the properties of the
dust, such as the dust temperature and the dust mass. To model the
dust emission from ARP~302N, we use the IR continuum fluxes measured
by Spitzer MIPS at 70 $\mu$m (5.16$\pm1.03$ Jy) and 160 $\mu$m
(5.08$\pm1.02$ Jy), and also by IRAS at 60 $\mu$m (5.32$\pm1.60$ Jy)
and 100 $\mu$m (11.6$\pm3.5$ Jy) (see details in \S\ref{spitzer})
together with the flux density of 135$\pm24$ mJy at 850 $\mu$m
reported by \cite{dunne00}.

\begin{figure}
  \includegraphics[angle=0,width=0.5\textwidth]{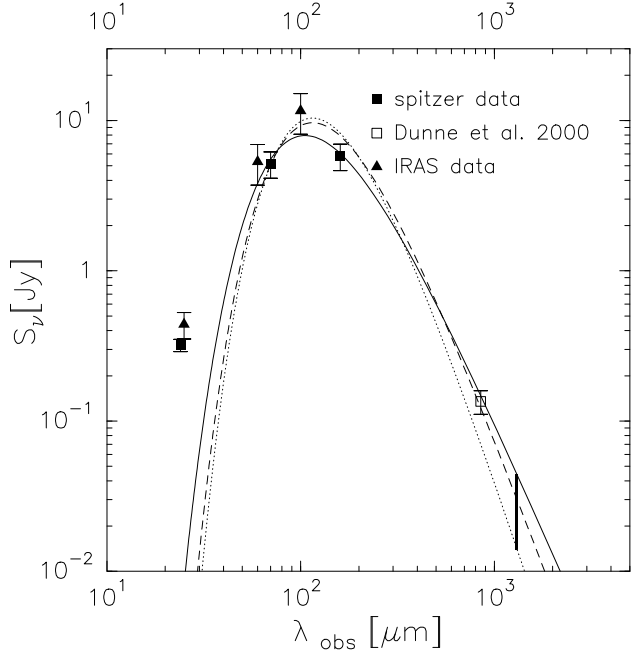}\\
  \caption{Observed SED of ARP~302N fitted by different single component dust models.
  The single component dust model are shown for ($\beta$, $\rm T_{dust}$, $\rm M_{dust}$)
  combinations as follows: the best fit (solid line: 1.0, 36 K, 2.2$\times 10^8$ M$_\odot$);
  moderate $\beta$ value (dashed line: 1.5, 29 K, 2.0$\times 10^8$ M$_\odot$); the simple
  modified blackbody fit (dotted line: 2.0, 26 K,
  1.1$\times 10^8$ M$_\odot$). The vetical line shows the predicted value range at
  1.3 mm. Note that 24 $\mu$m data are not used for the fitting.}\label{sed}
\end{figure}

Assuming that the dust emission is optically thin, the flux
density of the dust emission is related to the dust temperature
and the dust mass by the relation: $\rm
S_{\nu}=(1+z)[B_{\nu}(T_{dust})-B_{\nu}(T_{bg})]\kappa_d(\nu)
M_{dust}/D_L^2$, where $\rm S_{\nu}$ is the observed flux density,
$\rm T_{dust}$ and $\rm T_{bg}$ are the dust and background
temperature respectively, $\rm B_{\nu}$ is the Planck function,
$\rm D_L$ is the luminosity distance of the source, $\rm M_{dust}$
is the dust mass and $\rm \kappa_d$ is the dust absorption
coefficient. For the dust absorption coefficient, we adopt $\rm
\kappa_d(\nu) = 0.4~(\nu/250~GHz)^\beta$ cm$^2$ g$^{-1}$, where
$\beta$ is the emissivity index. We note that the emisivity index
$\beta$ depends on the dust properties and needs to be specified
beforehand. By fitting the observational measurements, we can
derive the dust mass and dust temperature for each specified value
of $\beta$. In Figure~\ref{sed}, we show three fitting results
with ($\beta$, $T\rm _{dust}$, M$\rm _{dust}$) combinations:
$\beta$=1.0, $T\rm _{dust}$=36 K and M$\rm
_{dust}$=2.2$\times$10$^8$ M$_\odot$ (the best fit), $\beta$=1.5,
$T\rm _{dust}$=29 K and M$\rm _{dust}$=2.0$\times$10$^8$ M$_\odot$
(moderate $\beta$ value), and $\beta$=2.0, $T\rm _{dust}$=26 K and
M$\rm _{dust}$=1.1$\times$10$^8$ M$_\odot$ (the simple modified
blackbody fit). The dust temperature ranges from 26 to 36~K for
the different models. Previous two models gives a dust mass of
2.0-2.2$\times$10$^8$ M$_\odot$, which are twice of the value from
the simple modified blackbody model.

The simple modified blackbody model underestimates
the 850~$\mu$m flux density by $\sim$60\%, which indicates that either
additional cold component or modifications to the emissivity index
are needed or that the optically thin assumption is inappropriate.
Actually, the gas is very extended in this galaxy and should result in
the similar distribution for the dust emission. The dust emission is 
still optically thin at 70~$\mu$m even if the dust is distribuated 
over a region with a small radius of 1~kpc, with a dust mass of 
2.0$\times$10$^8$ M$_\odot$,
thus the optically thin assumption should be correct in this galaxy.
In a large survey for local galaxies by \citet{dunne00}, the dust 
emissivity indexes to the best fitting models are found to be
1.3$\pm$0.2 and they argued that the galaxies may contain a
significant amount of cold dust. For ARP~302N, an additional cold gas component 
with a dust mass of 1.6--2.5$\times$10$^8$ M$_\odot$ and a dust temperature 
of 20-15~K can conpensate for the lower predicted flux at 850~$\mu$m. 
This results in a total
dust mass of 2.7--3.6$\times$10$^8$ M$_\odot$ dependent on the
temperature of the cold dust component.

The observed flux in CO $J$=1$-$0 \citep{lo97} is 304 Jy~\kms for ARP~302N,
and the corresponding CO line luminosities are 1.6$\times$10$^{10}$ K~\kms~pc$^2$.
This places ARP~302 to be the most CO luminous system known in the local
universe and possibly the most gas rich system. Usually, we can estimate
the gas mass using a standard conversion factor, 4.8~M$_\odot$ (K~\kms~pc$^2$)$^{-1}$
in our Galaxy, and it leads to a gas mass of 7.7$\times$10$^{10}$M$_\odot$.
We derived a dust mass of 2.0-3.6$\times$10$^8$ M$_\odot$ from our dust models as above.
Adopting the gas-to-dust mass ratio of 150 in our Galaxy, this yields
the value of 3.0-5.4$\times$10$^{10}$M$_\odot$ for the gas mass.
Therefore, the gas mass estimates from the dust and the CO luminosity
are comparable within their uncertainties and lead to a gas mass of
5.4$\times$10$^{10}$M$_\odot$ (the mean value from two methods).

The theoretical prediction of the flux density of the continuum
emission at 1.3mm is less than 42 mJy from the above dust models.
The sensitivity of the continuum observation is 5 mJy beam$^{-1}$,
thus it is reasonable that there is no positive detection from the
SMA observation if the emission is distributed over a region of
over 3 beams.

\subsection{Excitation condition of the gas}\label{excitation}

The excitation condition of molecular gas in ARP~302N can be
studied using the CO $J$=2$-$1/$J$=1$-$0 line ratio. As seen in
Figure~\ref{arp302_color} the line ratio peaks at either side of
the center. The ratio varies significantly across the galaxy, from
about 0.2 in the outer region to about 0.5 at the peaks. Assuming
that the missing flux of the SMA observation is uniformly
distributed, the line ratio will be 0.2 higher and the ratio will
vary from 0.4 in the outer region to about 0.7 at the peaks. The
overall line ratio from the single dish measurements is about 0.5.
From a survey of the CO $J$=1$-$0 and $J$=2$-$1 lines in the inner
kpc regions of a large sample of nearby spiral galaxies,
\cite{braine92} measured an average line ratio of 0.89, but
7 out of 35 galaxies have CO $J$=2$-$1/$J$=1$-$0 line ratios
less than 0.6. Furthermore, the gas in the central regions tend to 
have high excitation temperature, which will result in the higher 
line ratio if the CO line emission is optically thick.
The disks of NGC1808 \citep{aalto94} and our Galaxy show values
of 0.7 and 0.6, which are comparable to the value in the central
region of ARP~302N.

\begin{figure}
  \includegraphics[angle=0,width=0.9\textwidth]{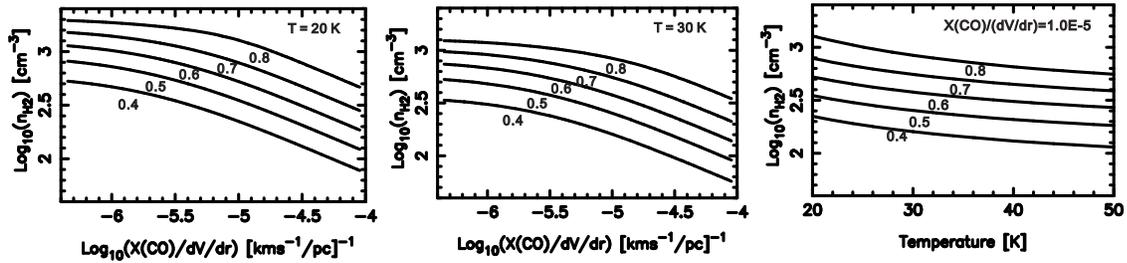}\\
  \caption{The dependence of CO $J$=2$-$1/$J$=1$-$0 line ratio on the
  gas density $n({\rm H}_2)$ and the abundance per velocity gradient
  [CO]/$\rm{(dv/dr)}$ for the kinetic temperature of 20~K and 30~K (two left panels), and
  the dependence of line ratio on the gas density and kinetic
  temperature T$_{\rm kin}$ for a fixed abundance per velocity
  gradient of [CO]/$\rm{(dv/dr)}$ = 1$\times$ 10$^{\rm -5}$ pc~($\rm
  km s^{-1}$)$^{-1}$ (right panel). Te solid lines show various values of 
  the line ratio between 0.4 and 0.8 by step of 0.1.}\label{lvg}
\end{figure}

To be more quantitative, we use our molecular excitation model
to constrain the physical conditions of the molecular gas. 
The transfer of radiation is calculated within the framework of
the large velocity gradient approximation. We take into account
all rotational levels of the CO molecule up J=10.
We use the collisional cross sections between CO and H$_2$ molecules 
from \citet{flower85}. The statistical equillibrium equations
are solved using the Newton-Raphson method. In our molecular excitation model 
three parameters are required: the gas density, the kinetic temperature and
the fractional abundance of the CO molecule per unit velocity gradient. 
At the expected low density of the molecular gas, the kinetic temperature 
is expected to decoupled from the dust temperature.
In this case, the kinetic temperature might be as low as half
of the dust temperature \citep{downes98}. In the previous section,
we estimate the dust temperature in the range of 26 to 36 K. For the
excitation calculation, we adopt two representative values of 20 and 30 K.
In Figure~\ref{lvg} (left panel) we present the dependence of CO
$J$=2$-$1/$J$=1$-$0 line ratio on the gas density $n({\rm H}_2)$
and the abundance per velocity gradient [CO]/$\rm{(dv/dr)}$ for 
kinetic temperature of 20 and 30~K. The figure shows that for the line ratio
measured in ARP~302N the gas density is low, less than 10$^3$
cm$^{-3}$. At this low density, the transition $J$=2$-$1 is
sub-thermally excited. Lower line ratios correspond to lower
densities for a reasonable range of abundance per velocity
gradient. To provide further support for this conclusion we also
show the dependence of line ratio on the gas density and kinetic
temperature T$_{\rm kin}$ for a fixed abundance per velocity
gradient of [CO]/$\rm{(dv/dr)}$ = 1$\times$ 10$^{\rm -5}$ pc~($\rm
km s^{-1}$)$^{-1}$ (right panel of Figure~\ref{lvg}). The result
suggests only a weak dependence on kinetic temperature. Our
finding is very similar to the conclusion of \cite{sakamoto94} in
that the variation in line ratio mainly reflects the variation in
gas density. To set more stringent limits on the gas density and
temperature would require observations of higher transitions of CO
and/or its $^{13}$CO isotope.

The relatively low gas density in ARP~302N is reflected
in its spatial distribution. For ULIRGs, which are in a more
advanced merger phase, \cite{downes98} show that the molecular gas
is more strongly concentrated in the nuclear region of the galaxy in
the form of a rotating disk with size of 1 kpc or less. The
estimated average gas density is higher, ranging from a few 10$^3$
to 10$^4$ cm$^{-3}$. In ARP~302N, there is a large amount of
molecular gas (5.4$\times$10$^{10}$ M$_{\odot}$, see in \S\ref{dust}), but this gas is
distributed over a very extended galactic disk with a size of 15
kpc, resulting in lower average density.
We note that the diffuse and very extended spatial distribution 
of molecular gas in ARP~302N is unusual among local luminous and ultralumious
galaxies. However, in some cases such as in the interacting galaxy M51, the molecular
gas traced by CO emission also covers all the optical disk, 
out to at least 12 kpc radius \citep{schuster07}. 
The origin of this distribution is currently unknown.

\subsection{Star forming activity}\label{formation}
The star formation activity can be estimated by SFE, the ratio of luminosity to molecular gas mass,
of the galaxies. For ARP~302N, we estimate an infrared luminosity
of 3.4$\times$10$^{11}$ L$_{\odot}$, separating the contamination
of ARP~302S in the manner described in \S\ref{spitzer}. We can
also estimate the infrared luminosity from the 24 $\mu$m data with
the empirical formula L$_{IR}$=10$\nu L_\nu(24 \mu m)$
\citep{bell05}, yielding a luminosity of 2.7 $\times$10$^{11}$
L$_{\odot}$, which is comparable with our estimated IRAS
luminosity. The molecular gas estimate is 5.4$\times$10$^{10}$ M$_{\odot}$ 
for ARP~302N. Its SFE of 6.3 L$_{\odot}$/M$_{\odot}$ is slightly 
higher than that in our Galaxy but much
lower than starburst galaxies or (U)LIRGs. The high infrared
luminosity, the large amount of the gas mass and the low SFE make
this galaxy peculiar.

The SFR for ARP~302N is about 40 $\rm
M_\odot~yr^{-1}$ assuming SFR = 1.17$\times$10$^{-10}$ ($\rm
L\rm_{FIR}/L_{\odot}$) $\rm M_\odot$ yr$^{-1}$
\citep{kennicutt98}, and thus ARP~302N can retain such a
luminosity over a period of at most 2.0$\times 10^9$ yrs. The
system will likely evolve into advanced merger in a timescale of
several tens million yrs \citep{mihos96}, and may at that time
become a ULIRG with a luminosity of 1.6$\times$10$^{12}$
$L_{\odot}$ even if it will lose half of its gas mass by then
(given a SFE of 42 L$_{\odot}$/M$_{\odot}$ for the merging systems
with the separation less than 5 kpc from Gao \& Solomon 1999).

\subsubsection{Constraint from the line ratio}
As discussed in previous section, our excitation analysis suggests
that the low line ratio in ARP~302N is due to the low volume density
of the molecular gas. The low gas density is entirely consistent
with the low SFE observed in ARP~302N, as star
formation is associated with dense gas rather than just total gas
mass \citep{gao04}.

\subsubsection{Radio continuum}
The thermal radio emission from HII regions is identifiable by its
flat spectral index with a typical value of 0.1. A spectral index
of 0.9 as measured for ARP~302N implied that non-thermal emission
dominates. Synchrotron emission with a spectral index of 0.9 is
likely to be dominated by supernova remnants with small diameters,
referred to the star forming activity in ARP~302N. From the high
resolution images at 3.6~cm, the continuum emission mainly comes
from the nucleus and ARP302N-cm. Therefore, the nucleus and
ARP302N-cm are the regions with strong star forming activity.

\subsubsection{Strong star forming activities at the nucleus and ARP302N-cm}

The extended infrared emission from Spitzer 8 and 24 $\mu$m data
excludes the existence of a strong AGN in ARP~302N and favors the
star formation origin for the large infrared luminosity in
ARP~302N. Therefore it fully supports that the star formation
dominates the IR emission in the galgaxy. The 3.6 cm observation
shows that there is strong and extended radio continuum emission
at ARP302N-cm. The Spitzer 8 and 24 $\mu$m images peak at the same
region, indicating that there is strong emission from the PAH and
hot dust. The high resolution CO $J$=2$-$1 image from the SMA
reveals much molecular gas exists at ARP302N-cm. All of these
results support that there is the asymmetrical star forming
activities in the system and ARP302N-cm is the strongest active
star forming region in the galaxy.

From 3.6~cm images with the different
resolutions at the nuclear region, it is clear that at least 36\%
of the emission is not associated with the unresolved structure in 
the nuclear region and 64\% of the emission is originated from a 
compact region (from the VLA 3.6~cm data, the deconvolved size is
0.23$\arcsec$$\times$0.09$\arcsec$, i.e. 143 pc $\times$ 56 pc). 
As a whole, the star formation is not active in ARP~302N, but there 
is strong star forming regions at the nucleus and ARP302N-cm.

From the previous optical spectroscopy observation
\citep{veilleux95}, the system was classified as LINER type galaxy. 
\citet{alonso00} claim that LINER-like optical 
spectra would be the consequence of the shock-heating by supernovae 
of the surrounding gas after a burst of star formation, as observed
in NGC~5218 \citep{olsson07}. This scenario to explain the LINER
type spectrum is consistent with our radio observation which suggests
that there are many supernova explosions. Of course, we can not
exclude the possible existence of a weak AGN in the center.

\subsection{Comparison with other mergers}\label{compar}
In Arp 302N the distribution and kinematics of molecular gas show no 
discernable disturbance. That suggests the Arp 302 system is in the 
early merging stage. Recently, many authors studied the molecular 
gas in high angular resolution to investigate 
the effects of the merging procedure and the star formation therein. 
The molecular gas is often concentrated on the overlap region or at least have 
another peak except for the centers of the systems with the well seperated nuclei, 
such as ARP~299 \citep{casoli99}, NGC~6090 \citep{wang04} and VV~114 \citep{iono2004}. 
In these cases, the starburst acitivites is prominent and the SFE is much higher 
than that in our Galaxies. NGC~6670 is another widely seperated merger 
and is well studied by \citet{wang01}. They also found that the stars 
and the molecular gas do not show any clear sign of the
merging process. However, the atomic gas is found to concentrate in the region in between
the two galaxies, a very clear signature of the merging process.
The atomic gas will be helpful to determine the merging process in ARP~302. 
Indeed, simulations show that the high SFRs and the SFEs arise only in the late 
stages of the merger and the star formation does not deplete much gas 
before it reaches the central regions in the mergering systems \citet{mihos94,mihos96}. 
It is consistent with the observed enhanced SFE in the 
late mergers by \citet{gao99}, which suggests that the fact before 
its final merging stage is the reason why the SFE in this LIRG is just comparable to the 
value in our Galaxy. The huge infrared luminosity in ARP~302N is just due to 
its large amount of the gas mass instead of the starburst activity. 

\section{Conclusion}\label{conclusion}
In this paper we present the SMA observations of the CO $J$=2$-$1
transition towards the northern galaxy, ARP~302N, of the early
merging system, ARP~302, to study the physical properties of the gas. 
To well understand the star formation activities,
the VLA data at 3.6~cm, 6.0~cm, 20~cm and the Spitzer (both IRAC
and MIPS data) archive data are also shown here. Our main results
are summarized as the following:

\begin{enumerate}

\item Our high angular resolution CO $J$=2$-$1 observation reveals
the extended spatial distribution of the molecular gas in
ARP~302N. We find that the molecular gas has a very asymmetric
distribution with two strong concentrations on either side of the
center together with a weaker one offset by about 8 kpc to the
north. This spatial distribution is very similar to the morphology seen
previously in the CO J=2--1 line.

\item The distribution of the molecular gas is also found to be
consistent with that for the PAH and the hot dust as traced by the 8 $\mu$m
and 24 $\mu$m continuum emission observed by Spitzer.

\item The line ratio CO J=2--1/1-0 shows that the line ratio
varies strongly from about 0.7 near the galaxy center to 0.4 in
the outer part of the galaxy. Excitation analysis suggests that
the gas density is low, less than 10$^3$ cm$^{-3}$, over the
entire galaxy.

\item By fitting the far infrared part of the SED of ARP~302N, we obtain a dust 
temperature $T\rm _d$ in the range 26--36 K and a dust mass of 
M$\rm _{dust}$=2.0--3.6$\times$10$^8$ M$\rm _\odot$.

\item We determine the spectral index of the radio continuum to be approximately 0.9. The
spatial distribution and spectral index of the radio continuum
emission suggests that most of the radio continuum emission is
synchrotron emission from the star forming regions at the nucleus
and ARP302N-cm. The good spatial correspondance between the 3.6 cm radio continuum
emission, the Spitzer 8 and 24 $\mu$m data and the high resolution
CO $J$=2$-$1 observation from the SMA shows that there is the
asymmetrical star forming activities in ARP~302N and ARP302N-cm is
the strongest active star forming region in the galaxy.

\end{enumerate}

\acknowledgments

We would like to acknowledge the anonymous referee for the valuable discussions and
comments. We also thank the SMA staff for maintaining the operation of the array.
The SMT is operated by the Arizona Radio Observatory (ARO), Steward
Observatory, University of Arizona. Y.-P. Ao thanks the ASIAA for
supporting his stay as a Visiting Scholar, when much of this work
was done. Y.-P. Ao also acknowledge the support of NSFC grant 10733030.

\begin{deluxetable}{cccccc}
\tablewidth{0pc} \tablecaption{Basic parameters for the SMA and
VLA observations towards ARP~302N \label{xband}} \tablecolumns{6}
\tablehead{
    \colhead{Data}    & \colhead{Freq.} & \colhead{Line/cont.} &  \colhead{beam size, P.A.}
& \colhead{Image rms} &  \colhead{Intensity\tablenotemark{a}}  \\
    \colhead{}    & \colhead{GHz} & \colhead{} &  \colhead{arcsec,degree}
& \colhead{mJy~beam$^{-1}$} &  \colhead{mJy\tablenotemark{b}}   }
\startdata
SMA   & 222.967 & CO $J$=2$-$1  & 3.4$\times$2.5, 154$\rm ^o$ & 60\tablenotemark{c}  & 382$\pm$10  \\
VLA L & 1.49 &  continuum   &  14$\times$13, 140$\rm ^o$ & 1.6  & 89.0$\pm$3.6 \\
VLA C & 4.86 &  continuum   &  15$\times$13, 178$\rm ^o$& 0.21 & 30.2$\pm$0.4 \\
VLA X (taper 150 k$\lambda$)  & 8.44 &  continuum   & 0.95$\times$0.90, 59$\rm ^o$   & 0.14 & 1.70$\pm$0.31;2.53$\pm$0.48 \\
VLA X (no taper)              & 8.44 &  continuum   & 0.24$\times$0.23, 87$\rm ^o$   & 0.05 & 1.08$\pm$0.12;0.93$\pm$0.22 \\

\enddata
\tablenotetext{a}{For VLA 3.6~cm data, the first column for the
nucleus and the second column for ARP~304N-cm.}
\tablenotetext{b}{For the SMA result, the integrated intensity is
in unit of Jy km~s$^{-1}$.} \tablenotetext{c}{Velocity
resolution of 20 km~s$^{-1}$.}

\end{deluxetable}

\end{document}